\documentclass[12pt,preprint]{aastex}  
\shorttitle{IC4296}
\shortauthors{Pellegrini et al.}

\received{2002 September 30}
\begin{document}

\title{The nuclear accretion in the FR I radio
 galaxy IC4296\\ from CHANDRA and VLBA observations }

\author{S. Pellegrini\altaffilmark{1}, T. Venturi\altaffilmark{2},
A. Comastri\altaffilmark{3}, G. Fabbiano\altaffilmark{4},
F. Fiore\altaffilmark{5}, C. Vignali\altaffilmark{6},
R. Morganti\altaffilmark{7}, G. Trinchieri\altaffilmark{8}}
\affil{{\rm $^1$Astronomy Department, Bologna University, via Ranzani
1, 40127 Bologna, Italy}} 
\affil{{\rm
$^2$Institute of Radioastronomy, CNR, Via Gobetti 101, Bologna,
Italy}} \affil{{\rm $^3$INAF--Bologna Astronomical Observatory, via
Ranzani 1, 40127 Bologna, Italy}} \affil{{\rm $^4$Center for
Astrophysics, 60 Garden Street, Cambridge, MA 02138, USA}} \affil{{\rm
$^5$INAF--Rome Astronomical Observatory, via Frascati 33, Monteporzio
Catone, Italy}} \affil{{\rm $^6$Dept. of Astron. and
Astrophys., PSU, 525 Davey Lab, Univ. Park, PA
16802, USA}}
\affil{{\rm $^7$NFRA, Postbus 2, AA 7990 Dwingeloo, The Netherlands}}
\affil{{\rm $^8$INAF--Brera Astronomical Observatory, via Brera 28, Milan, Italy}}

\begin{abstract}

A high angular resolution study of the nucleus of the FR I galaxy
IC4296 using $Chandra$ ACIS-S and VLBA observations is presented, with
the aim of studying the nature of the accretion process.  Pointlike
and hard X-ray emission is found, well described by a moderately
absorbed power law of $\Gamma=1.48^{+0.42}_{-0.34}$; no iron
fluorescence line from cold material is detected. The 0.3--10 keV
luminosity is $2.4\times 10^{41}$ erg s$^{-1}$, that is $\sim 400$
times lower than the accretion luminosity resulting from the estimated
Bondi mass accretion rate and a radiative efficiency of 10\%.  On the
parsec scale a jet and a counter-jet extend out from a central
unresolved ``core'' in the 8.4 GHz image.  Their orientation is in
good agreement with that of the large scale jets and their bulk speed
is relativistic. The parsec scale spectrum is convex over 2--22 GHz.
The observed nuclear luminosity is not likely to be reconciled
with the accretion luminosity by assuming that Compton thick material
surrounds the nucleus.  Low radiative efficiency accretion flow models
(ADAF and its variants) cannot account for the observed emission and
spectral shape of the nucleus in the radio band.  The power in the
jets accounts for a sizable fraction ($\ga 10$\%) of the accretion
luminosity; therefore the mass accretion rate needs not to be much
different from the estimate obtained using Bondi's theory.  A jet
dominated origin also for the observed radiative losses of the nucleus
is suggested.  This could also explain the finding that a nuclear
luminosity orders of magnitude lower than that of ``normal'' radio
loud AGNs is accompanied by a relatively higher radio emission.
 
\end{abstract}


\section{Introduction}

In recent years, increasing evidence has accumulated for a widespread
presence of central dark objects, most likely massive black holes
(MBHs), in nearby galaxies (Magorrian et al. 1998).  A possible
relationship between these galaxies and the relics of the ``quasar
era'' has therefore been suggested (Richstone et al. 1998).  Most
nearby galaxies, though, exhibit little or no nuclear activity in the
optical band (e.g., Kormendy and Ho 2000).  Also, their MBHs appear to
be starving. For example, early-type galaxies frequently host hot
gaseous halos that, if accreting onto the central MBHs, should give
rise to active nuclei with luminosities exceeding $10^{45}$ erg
s$^{-1}$ in standard accretion theories (Fabian and Canizares 1988; Di
Matteo et al. 2001a). Instead, recent $Chandra$ imaging of elliptical
galaxies with a massive hot interstellar medium showed nuclear
luminosities orders of magnitude lower than those expected from the
estimated Bondi mass accretion rates, for a radiatively efficient
accretion process (Loewenstein et al.  2001, Di Matteo et al. 2002).
Therefore, the question arises of how accretion proceeds at the
present time. This is also interesting in relation with the discovery
of a number of hard X-ray sources associated with otherwise ``normal''
early-type galaxies in Deep $Chandra$ surveys (Mushotzky et al. 2000,
Fiore et al. 2000). These could give a non negligible contribution to
the X-ray background (Barger et al. 2001).

In this paper the nature of accretion in the elliptical galaxy IC4296
is investigated, using $Chandra$ ACIS-S and VLBA observations. This
target is suitable for this study thanks to its proximity coupled with
the presence of a hot gaseous halo, very weak signs of activity in the
optical and an excess of hard X-ray emission over the expected
contribution from stellar sources, suggested by previous $ASCA$
observations (Sect. 2).  The combination of sensitivity, angular
resolution and spectral capabilities of ACIS-S allows us to determine
whether the excess of hard X-ray emission is nuclear in origin
and what its spectral properties and environmental conditions are. We
also conducted milliarcsec resolution VLBA observations of the nuclear
region to derive the radio emission properties on the parsec scale.
The X-ray and radio results are jointly used to shed light on the
accretion process.

In Sect. 2 we briefly describe the general properties of IC4296, 
in Sect. 3 we present the X-ray data analysis with particular 
emphasis on the nuclear emission, in Sect. 4 we present the results
of the VLBA observations, and in Sect. 5 we discuss the implications
of the observational results; in Sect. 6 our conclusions are summarized.

\section{Target properties}

IC4296 is a giant elliptical in the group of galaxies HG22 (Huchra and
Geller 1982; Table 1), at a distance of 49 Mpc.  Diffuse X-ray
emission at a temperature of $kT\sim 1$ keV around IC4296 had been
detected by the $ROSAT$ PSPC (Mulchaey et al. 1996).  $ASCA$ showed a
soft thermal component (of $kT\sim 0.8$ keV and abundance close to
solar at the best fit) coupled with a harder one, of roughly the same
fluxes in the 0.5--2 keV band (Matsumoto et al. 1997). The soft
component was better described by a cooling flow model rather than a
single temperature model, according to Buote and Fabian (1998).  The
harder component was modeled as bremsstrahlung emission with a
temperature ranging from 5 keV (Buote and Fabian 1998) to 10 keV
(Matsushita et al. 2000), or as a power law with photon index
$\Gamma=1.51^{+0.76}_{-0.20}$ (Sambruna et al.  1999). The latter
authors also place an upper limit of 740 eV on the equivalent width of
a 6.4 keV iron emission line.  Matsumoto et al. (1997) noted that the
ratio between the X-ray luminosity of the hard component and the
optical luminosity of the galaxy was higher than the average value
found for their sample of early-type galaxies.  Finally, IC4296 was
detected in the 5--10 keV band by the BeppoSAX High Energy Large Area
Survey (Fiore et al. 2001).

In the optical band, the continuum spectrum is typical of early-type
galaxies. No [O III] and weak [O II] emission lines were detected
(Wills et al. 2002).  A B--R color map shows large dust patches around
the center (Colbert et al. 2001). The galaxy was detected by $IRAS$ at
60 and 100 $\mu$m with flux densities presumably coming from cool
interstellar dust (Knapp et al. 1990).

IC4296 is also the optical counterpart of an extended FR I (Fanaroff
and Riley 1974) radio galaxy, PKS\,1333-33, whose total power is
logP$_{408 {\rm MHz}}$ (W Hz$^{-1}$) = 25.41 (Wright and Otrupcek
1990). VLA observations show a radio nucleus coincident with the
optical one, on the arcsecond scale, and two radio jets, almost
symmetric in brightness and extending in the sky for a total of
35$^{\prime}$ (Killeen et al. 1986).  A 2.3 GHz image
obtained with the southern hemisphere Long Baseline Array at the
resolution of a few tens of mas (i.e., a few parsecs) shows a
marginally resolved source with a total flux density on this scale of
$S_{2.3 {\rm GHZ}} = 128 \pm 12$ mJy (Venturi et al. 2000).  The radio
spectral properties of the nucleus are presented in Sect. 4 together
with the results of the high frequency VLBI observations conducted by us.

\section{X-ray observation and data analysis}

IC4296 was observed on 2001 December 15 with the ACIS-S camera
(Garmire et al. 2002) on board the $Chandra$ X-ray observatory
(Weisskopf et al. 2000).  The pointing placed the galaxy center in the
S3 chip and the total exposure was 24,835 s.  The 1/2 subarray mode
was used, to avoid possible pile-up. 
The present data analysis was done with the $Chandra$ analysis package
CIAO v2.2.1 and followed the $Chandra$ Data Analysis
Threads\footnote{http://cxc.harvard.edu/ciao/documents\_threads.html},
using the updated calibration files as of August 2001.  Level-2 event
files were considered.

Background flares are often seen in BI chips (S1 and S3). Therefore we
inspected the lightcurve of the exposed chip area after removal of
point sources, in the 2.5--7 keV band (which is most sensitive to the
most common flare species; Markevitch 2002). High background flares
were not found and the 
average count rate per square arcsec is consistent with that in the
blank field datasets (see the CIAO documents and
http://cxc.harvard.edu/contrib/maxim/acisbg/). It is also $\sim 2000$
times lower than that of a central circle of $2^{\prime\prime}$
radius (which is the extraction region used for the spectral analysis of 
the nucleus, Sect. 3.2).

\subsection{Image analysis}

The adaptively smoothed images of the central region of IC4296 are
shown in Figs. 1 and 2 for the 0.3--2 keV and 2--5 keV bands (the hard
band has been restricted to 5 keV to limit the background
contribution).  These images were produced using the $csmooth$ tool
with the S/N ratio set to be between 2.5 and 5.  Figs. 1 and 2 also
show the radio morphology of the source derived from a VLA image
obtained at 20 cm with a resolution of 3.2 arcsec (Killeen et
al. 1986).

The nuclear region is very bright both in the soft and in the hard
band. As can be judged also from the brightness profiles
(Figs. 3 and 4), in the central few arcsec the soft band emission is
much broader than that of the PSF, while an unresolved source
dominates in the hard band.  Diffuse hard emission appears only at
radii larger than a few arcsec, and is likely to be accounted for by
the unresolved X-ray binary population of the galaxy.  Much of the
soft emission in the central region likely comes from the 'cooling
flow' (Buote and Fabian 1998).  From Fig. 1 the emission seems to be
more elongated in the direction perpendicular to that of the radio
jets and no enhancement is seen coincident with the jet positions.  A
more detailed study of the morphology of the emission over the whole
chip area is in preparation.

A number of point sources are clearly detected; a study of the
properties of the X-ray point source population in the field is also
in preparation.  A wavelet detection algorithm (the $wavdetect$
program in CIAO) locates the X-ray center of the galaxy at
RA=13h36m39$^{s}$\hskip-0.1truecm .05,
Dec=--33$^{\circ}$57$^{\prime}$56$^{\prime\prime}$\hskip-0.1truecm .7,
in good agreement with the position of the radio center
(RA=13h36m39$^{s}$\hskip-0.1truecm .02,
Dec=--33$^{\circ}$57$^{\prime}$56$^{\prime\prime}$\hskip-0.1truecm .2
from the VLA image in Figs. 1 and 2).

\subsection{Spectral Analysis of the Nucleus}

The spectrum of the nuclear region was derived by extracting counts
from a central circle of $2^{\prime\prime}$ radius (95\% encircled
energy radius for 1.5 keV photons from a point source). The response
matrix and the effective area file were created using the mkrmf and
mkarf tools in the CIAO v.2.2.1 package with the N0002 (August 2001)
release of the response calibration files (FEF files).  The effective
area file was corrected
to account for the quantum efficiency degradation of ACIS at low energies
\footnote{See http://cxc.harvard.edu/cal/Links/Acis/acis/Cal\_prods/qeDeg/index.html and\\ http://www.astro.psu.edu/users/chartas/xcontdir/xcont.html.}.
Since the diffuse soft X-ray emission fills the entire S3 chip, the
background was estimated from the blank sky observations, 
for the same detector region considered for the
nucleus. This background contributes less than a count to the total
spectrum, in the 2.5--7 keV band.  In order to use the $\chi^2$
statistics, the data were grouped to include at least 20 counts per
spectral bin. Events with energies below 0.3 keV were excluded, since
the spectral calibrations are more uncertain at the lowest energies,
while the spectrum does not have enough source counts to be included
in the analysis above $\sim 7$ keV.  The results of the spectral
analysis are given in Table 2.

As expected, the spectrum is not well fitted by a single power law
(this leaves residuals below 1 keV) or a single thermal component
(this leaves residuals above 1 keV); in fact from the image analysis
the nuclear emission appears embedded within diffuse soft emission
(e.g., Fig. 1). An acceptable fit is obtained with two components: a
power law with photon index $\Gamma=1.48^{+0.42}_{-0.34}$, absorbed by
a column density $N_{H}=1.1^{+0.8}_{-0.5}\times 10^{22}$ cm$^{-2}$,
and soft thermal emission of $kT=0.56^{+0.03}_{-0.03}$ keV and $Z=1.0
Z_{\odot}$ (Figs. 5 and 6). The thermal emission could also be a
little absorbed (Fig. 7).  Some residuals appear between 0.7 and 1 keV
in Fig. 5; these are possibly due to line emission not accounted for
in a simple modeling of the thermal emission with a one temperature
plasma.  The 2--10 keV intrinsic luminosity of the hard component is
$1.6\times 10^{41}$ erg s$^{-1}$.  Replacing the power law component
with thermal bremsstrahlung produces a fit of the same quality; the
best fit temperature is $kT=26$ keV, with $kT>9$ keV at 90\%
confidence level.  A 6.4 keV (rest frame) Fe K$\alpha$ emission line
is not detected; the 90\% confidence upper limit on the equivalent
width of a narrow ($\sigma = 0.05$ keV) line at this energy is 460 eV.

\section{Radio observations}

PKS\,1333--33 was observed at milliarcsecond ($\sim 0.24$ parsec)
scale resolution with an array of 11 telescopes [the full VLBA (Very
Long Baseline Array) and one VLA antenna] on 13 February 2001.  The
observations were carried out simultaneously at 8.4 GHz, 15 GHz and 22
GHz, cycling among the three frequencies in order to ensure a good
u--v coverage. Given its low declination, the source was visible only
for $\sim$ 5 hours, so the total on source time was $\sim$ 1.5
hr/$\nu$.
The standard data reduction and image analysis were carried out using the
NRAO AIPS package.
Details on the observations, including the restoring beam (FWHM) and
noise level in the final images, are given in Table 3. 
Residual calibration errors are of the order of $\sim~$4\% at
each frequency.

PKS\,1333--33 is asymmetric on the parsec scale, as shown by the 8.4
GHz image in Fig. 8. At this frequency the source is characterised by
a central component, containing a flux density of 155 mJy ($\sim 90$\%
of the total on this scale), and two short jets. These are aligned
with a position angle of $\sim 140^{\circ}$; the orientation of the
parsec scale structure is in good agreement with that of the large
scale one, where the two jets show an average position angle of
$~\sim~130^{\circ}$ (Killeen et al. 1986).  The parsec scale jets
extend out to $\sim$ 5 mas (1.2 pc) from the central peak.
The northwestern jet is brighter and aligns 
with the brighter kiloparsec scale jet, as is usually found for
low-to-intermediate luminosity radio galaxies (e.g., Giovannini et al. 2001).
The source is unresolved at 15.4 GHz and 22.2 GHz.

\subsection{Parsec scale jet velocity}

If the asymmetric parsec scale morphology of PKS\,1333--33 is
attributed to Doppler boosting in an intrinsically symmetric source,
we can constrain the viewing angle $\theta$ and the bulk speed
($\beta=v/c$) of the jet.  At 1.2 pc from the peak, assumed to be the
core of the radio emission, the jet-to-counterjet brightness ratio is
$\sim 8$, which implies $\beta cos\theta~\sim~0.39$. This relation
leads to a maximum viewing angle $\theta_{max} = 67^{\circ}$, which is
consistent with the viewing angle estimated on the basis of the core
dominance ($\theta = 60^{\circ}$, Venturi et al. 2000). If we assume
that actually $\theta = 60^{\circ}$, we obtain that the jet is
relativistic on this scale, with a bulk speed $v = 0.79c$. This value
is in the range expected for a FR I radio galaxy (Giovannini et
al. 2001).

\subsection{The nuclear radio spectrum}

VLA observations show that the spectrum of the arcsecond
(kiloparsec) scale nucleus is convex: it is self-absorbed for $\nu
\leq $ 5 GHz and peaks at $\nu~\sim~15$ GHz (Killeen et al. 1986;
Morganti et al. 1997). The spectral index is $\alpha \sim -0.3$ for
$\nu~<~15$ GHz and steepens dramatically after the turnover, with
$\alpha \sim 1.8$ ($S_{\nu}\propto~\nu^{-\alpha}$).  In Fig. 9 the
radio spectrum of the kiloparsec scale nucleus is plotted, together 
with that of the parsec scale emission; the latter comes from the
flux densities derived here (Table 3) and the 2.3 GHz data given in
Venturi et al. (2000).  The parsec scale spectrum is
also convex: it is self absorbed up to $\nu~\sim 15$ GHz, with 
 $\alpha \sim -0.16$, and $\alpha \sim 0.25$ for $\nu > 15$ GHz.
From the parsec scale spectrum  a 2--22 GHz luminosity 
of $\sim 10^{40}$ erg s$^{-1}$ is derived.

There is a difference between the kiloparsec and parsec core flux
densities, and this steadily increases from 2.3 GHz to 15 GHz (the
ratio $S_{parsec}/S_{kiloparsec}$ goes from 0.75 to 0.58).  Given the
long time gap between the kiloparsec scale observations and ours, flux
density variability in the nucleus of PKS\,1333--33 might at least in
part be responsible for the discrepancy. However, the most
likely explanation is the existence of extended flux density resolved
out by our observations.

\section{Discussion}

The $Chandra$ ACIS-S observation of IC4296 reveals pointlike and hard
X-ray emission at the galaxy nucleus; this is well described by a
moderately absorbed ($N_{H}=1.1^{+0.8}_{-0.5}\times 10^{22}$
cm$^{-2}$) power law of $\Gamma=1.48^{+0.42}_{-0.34}$.  Its 0.3--10
keV luminosity  is $2.4\times 10^{41}$ erg s$^{-1}$
(Table 2). The 2--22 GHz
luminosity on the parsec scale is of the order of $\sim 10^{40}$
erg s$^{-1}$ (Sect. 4.2).
In the IR, only $IRAS$ data whose angular resolution
includes most of the galaxy body are available 
(Sect. 2); observations of the nucleus in the optical and UV bands are
lacking. If we conservatively assume a ratio $L_{{\rm bol}}/L_X \sim
10$ [the X-ray band has $\sim 30$\% of the bolometric luminosity in
Seyfert 1's, dropping to $\la 10$\% in the most luminous quasars,
Mushotzky et al. (1993)], then $L_{{\rm bol}}\sim 2.4\times 10^{42}$ erg
s$^{-1}$.  The correlation of MBH mass with central stellar velocity
dispersion of the host galaxy (Merritt and Ferrarese 2001, Gebhardt et
al. 2000) 
gives a black hole mass of $ 1.0\times 10^9$ M$_{\odot}$ for the
observed central stellar velocity dispersion of IC4296 (310 km s$^{-1}$,
Saglia et al. 1993).  From the numbers above, an Eddington ratio of
$L_{{\rm bol }}/L_{{\rm Edd}}\sim 2\times 10^{-5}$ is derived for the
nuclear emission of IC4296.

If a standard accretion disk is present, and the X-rays come from
Comptonization in a hot corona as devised for brighter AGNs (Haardt and
Maraschi 1991), this very sub-Eddington nuclear luminosity could be
explained by Compton thick material surrounding the nucleus and
heavily absorbing the X-ray emission. Then, similarly to what happens
in Seyfert 2 galaxies, only residual scattered emission is seen and a
strong 6.4 keV iron fluorescent line, of equivalent width $\ga 1$ keV,
is predicted (as in NGC1068; e.g., Matt et al.  1997).  This line has
not been detected in the nucleus of IC4296 and the upper limit on its
equivalent width is 460 eV (Sect. 3.2).  The $N_H$ value intrinsic to
IC4296 derived here is not very high ($N_H=1.1^{+0.8}_{-0.5}\times
10^{22}$ cm$^{-2}$) and consistent with that determined from a search
for HI absorption intrinsic to the IC4296 nucleus with the ATCA
(Morganti et al. 2001).
The presence of moderate absorption and, though at a lower confidence
level, the lack of a large equivalent width iron fluorescent line from
cold material are not new for FR I's nuclei: they have been found
already in a few cases studied with $Chandra$ (Hardcastle et al. 2002
and references therein; Chiaberge et al. 2002).  The suggestion that
thick tori are not present in FR I sources had also found support from
$HST$ observations of their optical nuclei (Chiaberge et al. 1999).

The low X-ray luminosity in presence of a supermassive black hole has
therefore to be explained in alternative ways.  One possibility is
that the mass supply rate to the central MBH is low. Alternatively, a
frequently suggested idea is that accretion occurs at low radiative
efficiencies, as in an advection dominated accretion flow (ADAF) and
its variants (e.g., Narayan 2002); this has been applied also to low
luminosity AGNs, ``normal'' ellipticals, and FR I galaxies (e.g., Di
Matteo et al. 2000, Di Matteo et al. 2001a). 
The other possibility would be that the central region of this galaxy
is not in a steady state, as for example in feedback modulated
accretion models (Binney and Tabor 1995, Ciotti and Ostriker 2001, Di
Matteo et al. 2002). In these models the interstellar medium is heated
by the impact of collimated outflows or by inverse Compton scattering
of hard photons, which stop or decrease recursively the accretion on
the central MBH.  Which of the above options applies is investigated
below with the help of our high resolution X-ray and radio
observations.

\subsection{Is the accretion rate low?}

The mass supply rate for accretion onto a black hole is usually
estimated adopting the Bondi accretion theory for spherical accretion
(Bondi 1952). This requires the calculation of the density and
temperature of the gas at ``infinity'' (in practice, near the
accretion radius $r_A$, defined as the point where the gravitational
potential of the MBH begins to dominate the dynamics of the hot gas;
$r_A\sim GM_{BH}/ 2c_s^2$, where $c_s$ is the sound speed). For
IC4296, $r_A\sim 100$ pc, corresponding to $\sim
0.42$ arcsec.  With ACIS-S data we can therefore derive the gas
distribution very close to the accretion radius.

The spectral analysis of a central $2^{\prime\prime}$ radius region
shows the presence of a hard component in addition to soft thermal
emission from the surrounding hot gaseous halo. Therefore we modeled
the central part of the 0.3--2 keV surface brightness profile (of
$\sim 20^{\prime\prime}$ radius, Fig. 3) as unresolved nuclear
emission, with a shape following that of the PSF for 2 keV photons,
plus a $\beta$-model profile ($\propto [1+(R/R_c)^2]^{0.5-
3\beta}$). The $\beta$-model flattens within $\sim 2^{\prime\prime}$,
where the nuclear point source dominates, and has $\beta=0.7$ at the
best fit.  The number of counts within a radius of $2^{\prime\prime}$
given by this profile decomposition for the unresolved nuclear
emission and for the $\beta$-model agrees with that resulting from the
spectral analysis for the hard component and the soft thermal emission
respectively.  The gas density profile is derived by deprojection of
the $\beta$-model used for the decomposition of the surface brightness
profile (e.g., Ettori 2000). Adopting the cooling function value for
the hot gas temperature and abundance derived from the spectral
analysis, the resulting central gas density is $\sim 10^{-24}$ g
cm$^{-3}$. From the Bondi (1952) analysis, the mass accretion rate
cannot exceed the value $\dot M_{Bondi}=4\pi \lambda_c (G\,M_{BH})^2 c_s^{-3}
\rho_{\infty}$, with $\lambda_c=0.25$ in the adiabatic case and
$\rho_{\infty}$ the gas density at ``infinity'' [see eq. (19) of Bondi
1952].  The value $\dot M_{Bondi}\sim 0.02 M_{\odot}$ yr$^{-1}$ is
then derived, when using the best fit gas temperature $kT=0.56$ keV
(Sect. 3.2).  This corresponds to an accretion luminosity
$L_{acc}=0.1\dot M_{Bondi}c^2\sim 10^{44}$ erg s$^{-1}$, for a
canonical radiative efficiency of 10\%, as assumed in a standard,
radiatively efficient thin accretion disk.  Therefore, radiative
losses from the nuclear region account for just $\la 2$\% of
$L_{acc}$. We must conclude that either accretion is indeed
stationary but takes place with a low radiative efficiency
(Sect. 5.2), and possibly there is another process into which a
significant fraction of $L_{acc}$ is channeled (Sect. 5.3), or the
galaxy center is out of equilibrium and Bondi's theory does not apply.

\subsection{Is the radiative efficiency low?}

We explore here the possibility that much of the energy in the
accretion flow cannot be radiated and is carried through the event
horizon of the MBH, as in an ADAF.  The Eddington ratio of the nucleus
of IC4296 lies well within the required value for ADAFs ($L_{{\rm
bol}}/L_{{\rm Edd}} <10^{-2}$).  In these models the majority of the
observable emission is in the X-ray and in the radio bands (e.g.,
Narayan 2002). In the X-rays it comes from thermal bremsstrahlung with
a flat power law ($\Gamma < 1.5$ in the $Chandra$ band) or inverse
Compton scattering of soft synchrotron photons by the flow electrons
(this produces a steeper X-ray spectral shape).  In the radio band,
synchrotron emission arises from the strong magnetic field in the
inner parts of the accretion flow. The self-absorbed synchrotron
spectrum rises up to high radio frequencies, above which it abruptly
drops.  If a wind develops from the inner regions the synchrotron
emission in the radio band and the Compton emission in the X-ray band
are drastically reduced.

We can compare the case of IC4296 to those of NGC1399, NGC6166, M87,
for which ADAF modeling has been attempted recently (Loewenstein et
al. 2001; Di Matteo et al. 2001b, 2002). Like IC4296 these are low
power (FR I type) radio galaxies all with nuclear X-ray luminosities
$\sim 10^3$ times lower than the estimated accretion luminosity,
as determined from $Chandra$ observations.
Pure inflow ADAFs accreting close to the Bondi rate with a low
radiative efficiency have been constructed and compared to the overall
spectral energy distribution.  The models are normalized using the
observed X-ray luminosities, while the two major parameters on which
the predicted spectrum depends are the accretion rate and the MBH mass
(e.g., Di Matteo et al. 2002).  The parameters of NGC6166
are close to those of IC4296 ($\dot M_{Bondi}\la 0.03 
M_{\odot}$ yr$^{-1}$ and $M_{BH}=10^9 M_{\odot}$).  A
pure inflow ADAF for NGC6166 has $\nu L_{\nu}($2 keV)$/\nu
L_{\nu}($22 GHz)$\sim 100$ (Di Matteo et al. 2001b), while this ratio
is $\sim 4$ for IC4296 (Fig. 10).  
Therefore the X/radio luminosity ratio of the IC4296 nucleus is much
lower than in this model.
Note that the radio emission from
an ADAF cannot be increased by appealing to variants including
outflows or convection, because these suppress the radio emission with
respect to the X-ray one (Quataert and Narayan 1999).  Therefore the
presence of an accretion flow with low radiative efficiency cannot be
excluded, but most of the radio emission should come from another
source. 
The same conclusion was reached also for NGC6166, due to its 
high VLBI emission (similar to that of IC4296, Fig. 10).
In the case of M87, $\dot M_{Bondi}=0.1 M_{\odot}$ yr$^{-1}$ and
$M_{BH}=3\times 10^9$. Again it
was found that pure inflow ADAFs can explain the observed X-ray
emission, but the majority of the radio emission must be accounted for
by a source different from the accretion flow (Di Matteo et al. 2002).

In addition to being too radio loud for ADAF models, the nucleus of
IC4296 has also a different spectral shape in the radio band: while in
pure inflow ADAFs the radio emission continues to rise at frequencies
larger than 22 GHz, since the radio spectral index between 1--100 GHz
is inverted with $\alpha \sim -0.4$ (Yi and Boughn 1999), the nucleus
of IC4296 turns over after 15 GHz (Fig. 9).  This turn over is shown
also by NGC1399, whose radio emission peaks between $\sim 10 - 30 $
GHz (this feature is visible in Fig. 10 if luminosities are divided by
frequencies).  This turn over is typical also of the nuclei of other
few ellipticals, whose nuclear radio emission was best modeled by
self absorbed synchrotron emission from small-scale jets (Di Matteo et
al. 2001a).

\subsection{Jet dominated energy budget}

We examine here the possibility that energy is carried off in an
outflow.  In fact, two jets are visible on the parsec scale
(Sect. 4).  Does the jet power amount to a sizable fraction of the
accretion luminosity $L_{acc}$ estimated in Sect. 5.1?

The total energy flux carried by a jet can be estimated as $P_j\sim
4\pi r_j^2 v_j \gamma_j^2 p_j$, with $r_j,p_j,\gamma_j,v_j$ being the
jet radius, pressure, Lorentz factor, and velocity (e.g., Owen et
al. 2000). We can derive the jet minimum power $P_{j,min}$ by
inserting the minimum pressure $p_{j,min}$ derived specifically for
the jets in IC4296: $p_{j,min}= 1.5\times 10^{-11}$ dyne cm$^{-2}$ at a
distance of $20^{\prime\prime}$, where the jet radius is $
5^{\prime\prime}$ and the jet velocity $v_j=2\times 10^4$ km s$^{-1}$
(Killeen and Bicknell 1988). Then $P_{j,min}= 5.2\times 10^{42}$ erg
s$^{-1}$, that is $\sim 10^{43}$ erg s$^{-1}$ for two jets. Therefore
$P_{j,min}\sim 0.1 L_{acc}$.  $P_{j,min}$ could be larger if the jet
velocity at $20^{\prime\prime}$ is higher, which depends on the age of
the source (the canonical age of $10^8$ yrs has been assumed by
Killeen and Bicknell, who leave open the possibility for the
source to be younger).  Given also that the calculation above gives
the {\it minimum} power of the jets, and that plausibly the jets have spent
energy (e.g., doing work against the interstellar medium)
 during their propagation from the nuclear region to a distance
of $20^{\prime\prime}$, we consider safe to conclude that the power
required by the jets at the parsec scale is $\ga 10$\% of $L_{acc}$.
Therefore the mass supply rate to the MBH needs not
to be much lower than $\dot M_{Bondi}$.

A correlation has been found between jet kinetic power and narrow line
luminosity for radio sources (Rawlings and Saunders 1991). When
inserting the line luminosities observed for IC4296
[log$F_{[OIII]}$(erg cm$^{-2}$ s$^{-1})<-14.43$, log$F_{[OII]}$(erg
cm$^{-2}$ s$^{-1})=-14.09$, Wills et al. 2002], the resulting jet
kinetic power is $\sim $ few$\times 10^{42}$ erg s$^{-1}$, similar to (but
somewhat lower than) the $P_{j,min}$ value estimated above (see
Sect. 6 for a caveat concerning possible underluminosity in line
emission of the nucleus of IC4296).  

Note that the jet could dump energy into the ISM
and stifle the accretion in the innermost part, thus reducing the
accretion rate with respect to $\dot M_{Bondi}$, as described by Di Matteo 
et al. (2002) for the nucleus of M87. In that case the estimated
decrease in the mass accretion rate is more than enough to relax the
requirement for low radiative efficiency of the accreting gas.

\subsubsection {Jet dominated nuclear emission}

We suggest that a large fraction of the nuclear radiative losses are
also due to the very inner jets. This seems reasonable for the
parsec--scale radio emission: its spectral shape (Sect. 4.2) can be
explained with nonthermal synchrotron emission from the base of a
self-absorbed jet, as is typical of jet models for powerful AGNs
(e.g., Zensus 1997). A jet origin for the nuclear X-ray emission is
consistent with its spectral shape of a power law, similar to that
found in the radio band ($\Gamma=1.48^{+0.42}_{-0.34}$ in the X-rays
and $\Gamma \sim 1.25$ for $\nu > 15$ GHz).  If we extrapolate the
optically thin part of the radio synchrotron spectrum of IC4296 to the
X-ray band, we obtain an X-ray emission much higher than observed.
Therefore, in a jet dominated model the spectral index should steepen
between the radio and X-ray frequencies and inverse Compton processes
(such as synchrotron self-Compton) should account for the X-ray
emission.  For example, a jet dominated model has been elaborated
recently to explain the spectral energy distribution of low luminosity
AGNs (i.e., sources with $L_X\la $few$\times 10^{41}$ erg s$^{-1}$;
see, e.g., Yuan et al. 2002 for the modeling of NGC4258 and Fabbiano
et al. 2002 for that of IC1459).  In this model
the X-ray spectrum is produced by  self-Comptonized emission of the
radio jet plus an extrapolation of the radio synchrotron emission. 
IC1459 is, similarly to IC4296, a bright radio elliptical 
with a nuclear X-ray emission $\sim
10^{4}$ times lower than the estimated accretion luminosity, from a
$Chandra$ pointing (Fabbiano et al. 2002). In the model elaborated for
its nucleus, the jets are by far the dominant sink of power: they
carry $\sim 800$ times the energy that is radiated by the nucleus in
the X-ray band, as kinetic and internal energy.

Previous studies of low power radio galaxies with $Einstein$ and 
$ROSAT$ showed a correlation between the X-ray luminosity of a
nuclear unresolved component and that of the radio core, giving
support to a jet-related origin for at least some of the nuclear X-ray
emission (Fabbiano et al. 1984, Canosa et al. 1999).  
The few $Chandra$ observations of FR I sources performed so far reveal
the presence of a nuclear power law component of $\Gamma\sim 1.4-2$ at
the best fit, consistent with both an inverse-Compton and a
synchrotron origin for the X-ray emission (e.g., Hardcastle et
al. 2002 and references therein).
A synchrotron origin has been suggested also for the nuclear 
emission of FR I sources in the optical band, since $HST$ images revealed a
correlation between the nuclear optical luminosity and that of the
radio cores (Chiaberge et al. 1999).

\section{Conclusions and final remarks}

$Chandra$ ACIS-S and VLBA observations of the nucleus of the 
FR I galaxy IC4296 have been analyzed, with the aim of studying the 
nuclear accretion process. The following results have been obtained:

$\bullet$ hard and pointlike emission has been detected at the galaxy
center, that can be described by a power law of
$\Gamma=1.48^{+0.42}_{-0.34}$, only moderately obscured
($N_{H}=1.1^{+0.8}_{-0.5}\times 10^{22}$ cm$^{-2}$), with 2--10 keV
luminosity of $1.6\times 10^{41}$ erg s$^{-1}$.  A 6.4 keV (rest
frame) emission line is not detected and the upper limit on its
equivalent width is 460 eV. The bolometric luminosity of this
unresolved source is $\la 2\times 10^{-5}$ of the Eddington luminosity
for a $10^9 M_{\odot}$ MBH (that is supposed to be present at the center
of IC4296).

$\bullet $ the nuclear region is also bright in the soft band, 
with a radial profile much broader than that of the PSF.  Most of this
soft emission comes from hot gas at a temperature of
$kT=0.56^{+0.03}_{-0.03}$ keV and solar abundance (in a simple
modeling with a one temperature plasma).

$\bullet$ the 8.4 GHz image shows on the parsec scale a jet and a
counter-jet extending out from a central unresolved ``core'', with an
orientation that is in good agreement with that of the large scale
jets.  The northwestern jet is brighter; from Doppler boosting in an
intrinsically symmetric source, we constrain the jet viewing angle and
the bulk speed, which turns out to be relativistic ($v = 0.79c$). Both
are found in the range expected for a FR I radio galaxy.

$\bullet$ the parsec scale spectrum is convex: it is self absorbed up
to $\nu~\sim 15$ GHz, with spectral index $\alpha \sim -0.16$, and it
has $\alpha \sim 0.25$ for $\nu > 15$ GHz. On this scale the radio
luminosity is $L\sim 10^{40}$ erg s$^{-1}$.

$\bullet$ Compton thick material surrounding the nucleus is not likely
to explain the very sub-Eddington nuclear luminosity observed, because
a strong 6.4 keV iron fluorescent line has not been detected.  The
moderate value of the hydrogen column density intrinsic to IC4296
derived here is consistent with that determined from a search for
nuclear HI absorption with the ATCA.

$\bullet$ the mass accretion rate close to the accretion radius has
been derived using Bondi's spherical accretion formula. This is $\sim
0.02 M_{\odot}$ yr$^{-1}$ and corresponds to an accretion luminosity of
$\sim 10^{44}$ erg s$^{-1}$ for a radiative efficiency of
10\%, as assumed in a standard, radiatively efficient
thin accretion disk.  Therefore, radiation losses from the nucleus
account for $\la 2$\% of this estimate of the accretion luminosity.

$\bullet$ the high radio/X luminosity ratio and the spectral shape in
the radio band resulting from our observations are not consistent with
the predictions of accretion flow models with low radiative efficiency
as ADAFs and their variants. These flows cannot be
the only component of the system.

$\bullet$ the large scale radio jets require a power input that is a
significant fraction $(\ga 10$\%) of the estimated accretion luminosity.
Therefore it is likely that a power comparable to this luminosity is
channeled into the nuclear jets observed on the VLBA scale, that are
the base of the large scale jets. The mass accretion rate needs not to
be much different from the estimate obtained using Bondi's theory.

$\bullet$ the alternative possibility to account for a discrepancy
between the accretion luminosity and the observed power output would
be that the system is out of equilibrium, therefore Bondi's theory
does not apply.  This is the case described by feedback modulated
accretion flow models where the central engine undergoes on-off
activity cycles. 

$\bullet$ it is suggested that the radiative losses from the nucleus
are also jet dominated. The radio emission can be ascribed to
synchrotron emission from the base of a self-absorbed jet and
the X-ray emission may have a synchrotron self-Compton origin.

$\bullet$ nuclear luminosities orders of magnitude lower than the
estimated accretion luminosity and the likely absence of an obscuring
torus have been found recently in a few other FR I's nuclei observed
with $Chandra$ and with $HST$.

The results above suggest some final considerations. 
This nucleus is orders of magnitude less luminous than that of
``normal'' radio loud AGNs, but is relatively more powerful in the
radio band, as revealed by the comparison of the respective spectral
energy distributions (Fig. 10).  If we assume that the nuclear radio
luminosity of IC4296 is jet dominated, then this finding could be the
result of a larger importance of the jet in the nuclear radiation
losses.  In addition, bright AGNs follow correlations between nuclear
X-ray luminosity and emission line luminosities, which are explained
by photoionization of the gas surrounding the nucleus.  Is this source
as efficient as bright AGNs in producing line emission?  In the
extension to low luminosities of the correlation between [OIII] line
emission and $L(2-10)$ keV for radio loud AGNs (Sambruna et al. 1999),
the nucleus of IC4296 turns out to be underluminous in [OIII] for the
2--10 keV luminosity measured here. The power required to sustain the
jets in IC4296 could then be underestimated using the Rawlings and
Saunders (1991) relation, as done in Sect. 5.3. In conclusion, there
is clear evidence that in IC4296 the central engine works in a
different way than in radio loud AGNs.

\acknowledgments 
We thank the referee for useful comments, G. Brunetti and L. Ciotti
for useful discussions and N. Killeen for kindly providing the radio
data for Figs. 1 and 2.  S.P. acknowledges funding from ASI (contract
I/R/037/01) and MURST (contract Cofin 2000,
prot. MM02438375). A.C. acknowledges funding from ASI (contract
I/R/113/01) and Cofin 00-02-036.

\clearpage

\begin{table*}
\begin{center}
\caption{IC4296 properties.\label{tbl-1}}
\begin{tabular}{  l l l l l l l l l l }
\tableline\tableline
Type  & {RA\tablenotemark{a}} & {Dec\tablenotemark{a}} & {z\tablenotemark{b}}
 & {D\tablenotemark{b}} & {$R_e$\tablenotemark{a}}  &  {$B_0^T$\tablenotemark{a}} & Log$L_B$    &  {$N_H$\tablenotemark{c}}\\
      &(J2000)&J(2000)   &   &  (Mpc) & (arcsec) &  (mag)    &($L_{\odot}$) & (cm$^{-2}$) \\
\tableline
E0    & 13h36m39$^{s}$\hskip-0.1truecm.37 & -33$^{\circ}$57$^{\prime}$59$^{\prime\prime}$ & 0.012465 & 49 & 41 & 11.42 & 11.0    & $4.3\times 10^{20}$\\  
\tableline
\end{tabular} 
\tablenotetext{{\rm a}}{From de Vaucouleurs et al. (1991). $R_e$ is the effective radius.}
\tablenotetext{{\rm b}}{The redshift z is from Smith et al. (2000).
D is the distance derived with the surface brightness fluctuations method
(Mei, Silva \& Quinn 2000).}
\tablenotetext{{\rm c}}{From Stark et al. (1992).}
\end{center}
\end{table*}

\begin{table}
\begin{center}
\caption{Spectral results for the nuclear ($R<2^{\prime\prime}$) emission, fitted with a power law $+$ mekal ($\chi^2/\nu=91/66$).\label{tbl-2}}
\smallskip
\begin{tabular}{  l l l l l l l l l l l l l l }
\tableline\tableline
pow :  & \\
\tableline
$N_H(10^{22}$ cm$^{-2}$)            & 1.1 (0.6--1.9)   \\
$\Gamma$                     & 1.48 (1.14--1.90) \\
F$_{0.3-2 {\rm keV}}$(erg cm$^{-2}$ s$^{-1}$) & 4.5$\times 10^{-14}$ \\
F$_{2-10 {\rm keV}}$(erg cm$^{-2}$ s$^{-1}$)  & 5.1$\times 10^{-13}$ \\
L$_{0.3-2 {\rm keV}}$($10^{40}$  erg s$^{-1}$)         & 7.7 \\
L$_{2-10 {\rm keV}}$($10^{40}$  erg s$^{-1}$)         & 15.9 \\
\tableline
mekal : & \\
\tableline
$N_H(10^{22}$ cm$^{-2}$)            & 0.06 (0.01--0.15)\\
$kT$(keV)                              & 0.56 (0.53--0.59)\\
$Z(Z_{\odot})$                         & 1.0 ($>0.4$)      \\
F$_{0.3-2 {\rm keV}}$ (erg cm$^{-2}$ s$^{-1}$) & 1.7$\times 10^{-13}$ \\
F$_{2-10 {\rm keV}}$ (erg cm$^{-2}$ s$^{-1}$)  & 3.8$\times 10^{-15}$ \\
L$_{0.3-2 keV}$($10^{40}$  erg s$^{-1}$)         & 7.5 \\
L$_{2-10 keV}$($10^{40}$  erg s$^{-1}$)         & 0.1 \\
\tableline
\end{tabular}
\tablecomments{
The absorbing column $N_H$ is in addition to the Galactic
value in Table 1.
Values in parentheses give the 90\% confidence interval for one interesting
parameter. Fluxes are observed and luminosities are intrinsic (corrected
for absorption).}
\end{center}
\end{table}

\begin{table}
\begin{center}
\caption{The VLBA observation log.\label{tbl-3}}
\smallskip
\begin{tabular}{rlcc}
\tableline\tableline
Freq. & FWHM      & rms   & S$_{tot}$ \\
 GHz  & mas, p.a. & mJy/b &   mJy     \\
\tableline
8.4   & 5.0$\times$2.0, 45$^{\circ}$ & 0.068 & 173.5 \\ 
15.4  & 5.2$\times$1.3, $-6^{\circ}$ & 0.25  & 174.0 \\ 
22.2  & 5.0$\times$2.0, 0$^{\circ}$ & 1.4   & 157.9 \\ 
\tableline
\end{tabular}
\end{center}
\end{table}

\clearpage

\begin{figure}
\plotone{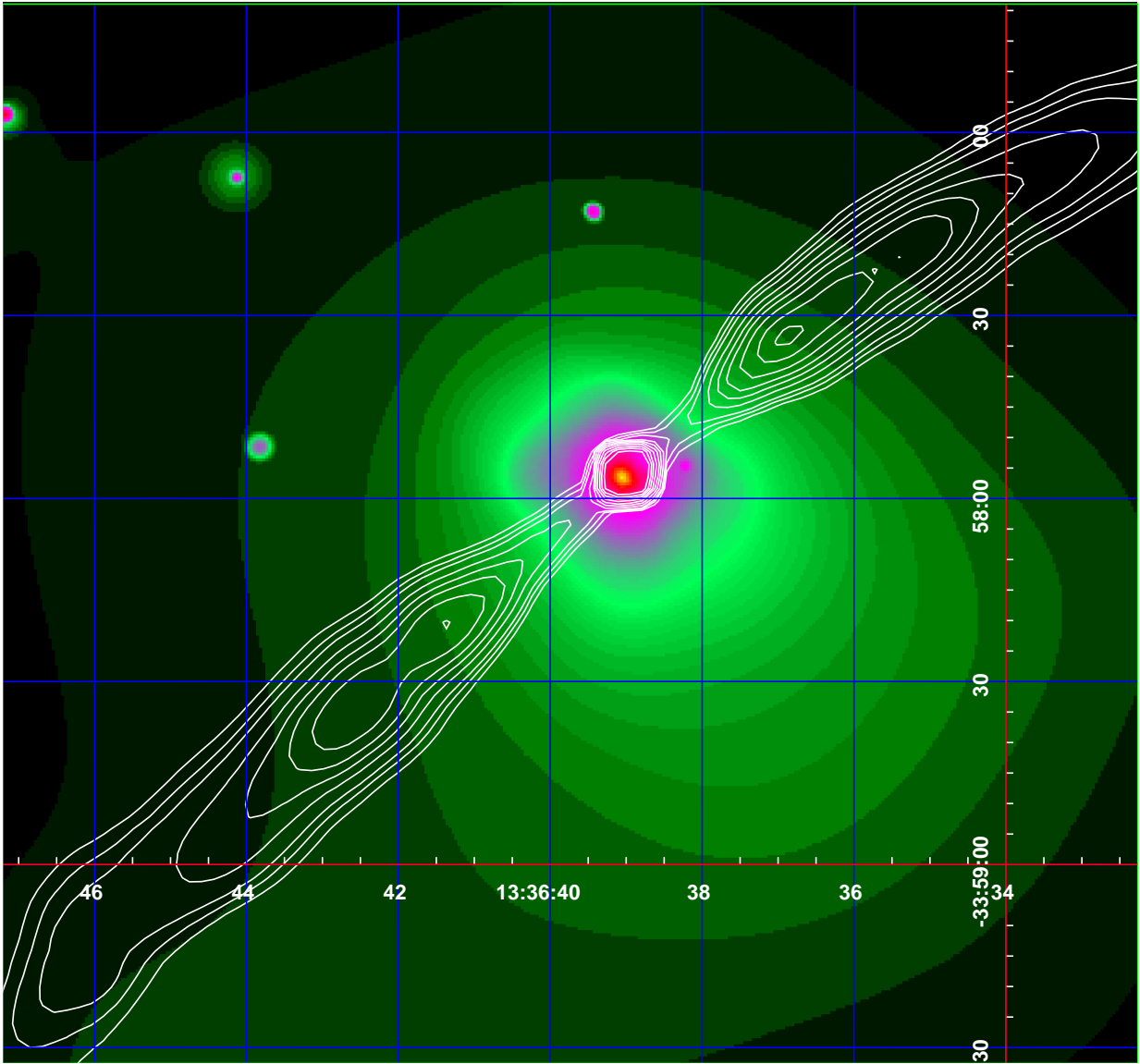}
\caption{Adaptively smoothed and
exposure corrected 0.3--2 keV image of the $3^{\prime}\times 3^{\prime}$ field
around the center of IC4296 (Sect. 3.1).  The color scale
is logarithmic and covers surface brightness levels that range over a
factor of $\sim 5000$.  The white contours show the morphology of the
radio source from a 20cm VLA image with a 3.2 arcsec FWHM beam; peak flux is
153.1 mJy per beam and the contour levels are -1\%, 1\%, 2\%, 3\%,
4\%, 5\%, 7\%, 8\%, 9.5\%, 11.5\% and 50\% of the peak (Killeen et al.
 1986). \label{fig1}}
\end{figure}

\clearpage 

\begin{figure}
\plotone{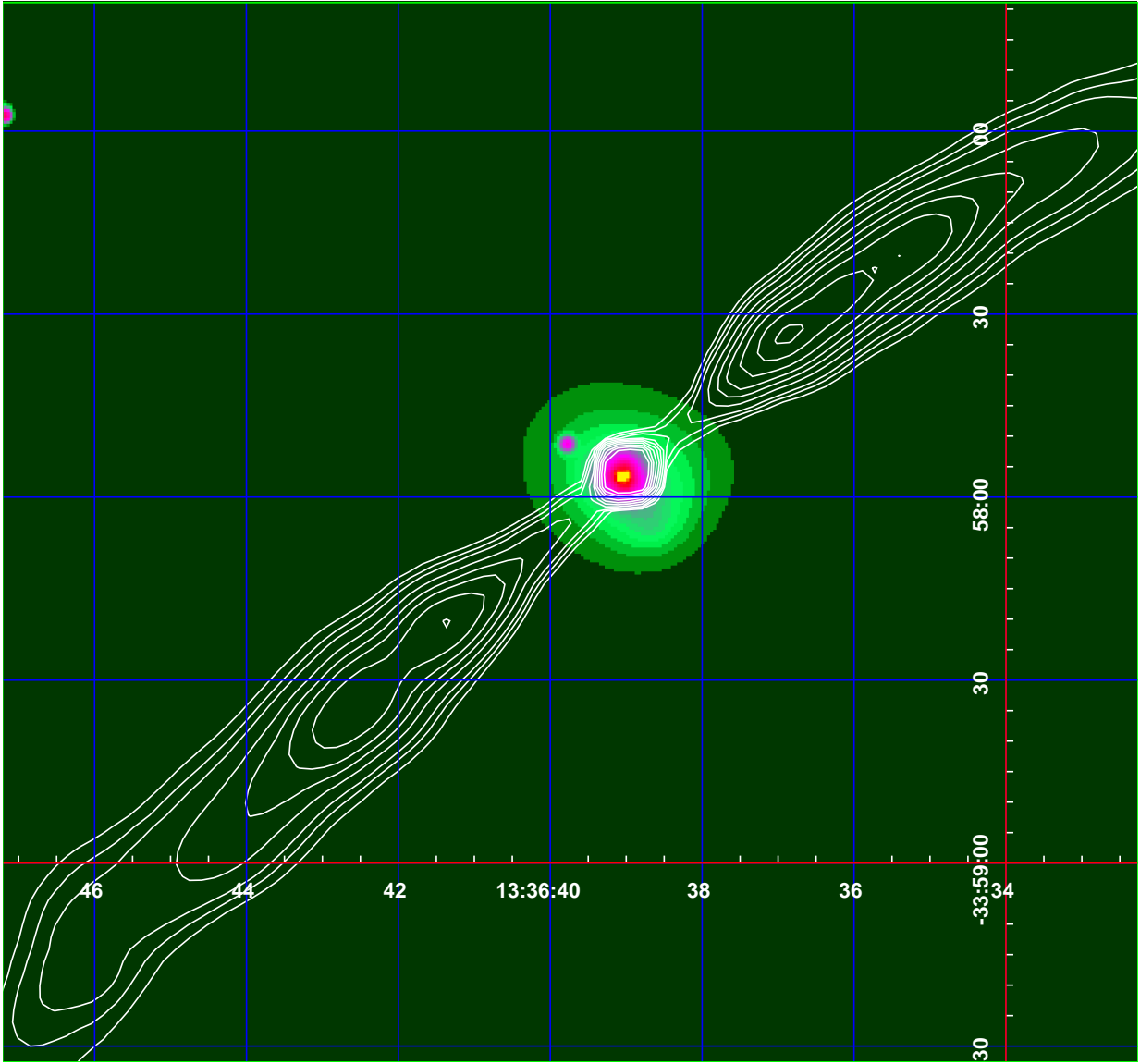}
\caption{Adaptively smoothed 2--5 keV image of the same field as in
Fig. 1, together with the same radio contours.  The color scale covers
surface brightness levels that range over a factor of $\sim
3000$.\label{fig2}}
\end{figure}

\clearpage

\begin{figure}
\plotone{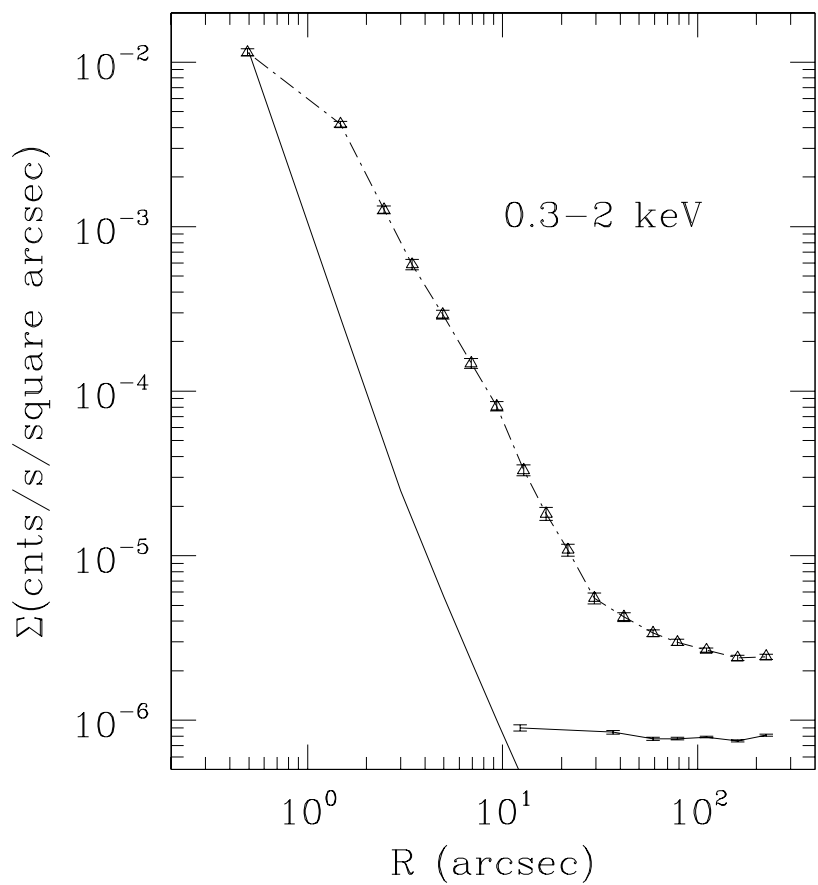}
\caption{The 0.3--2 keV surface
brightness profile of the total (source+background, triangles) and
background (solid line) emission around the center of IC4296. The
background has been estimated from the blank sky observations.
The PSF profile for 2 keV photons is shown with a solid line. It has
been estimated from the standard ACIS-S PSF library files for the
source location in the telescope field of view. Point sources detected
in the field have been removed from the calculation of the
profile.\label{fig3}}
\end{figure}

\clearpage

\begin{figure}
\plotone{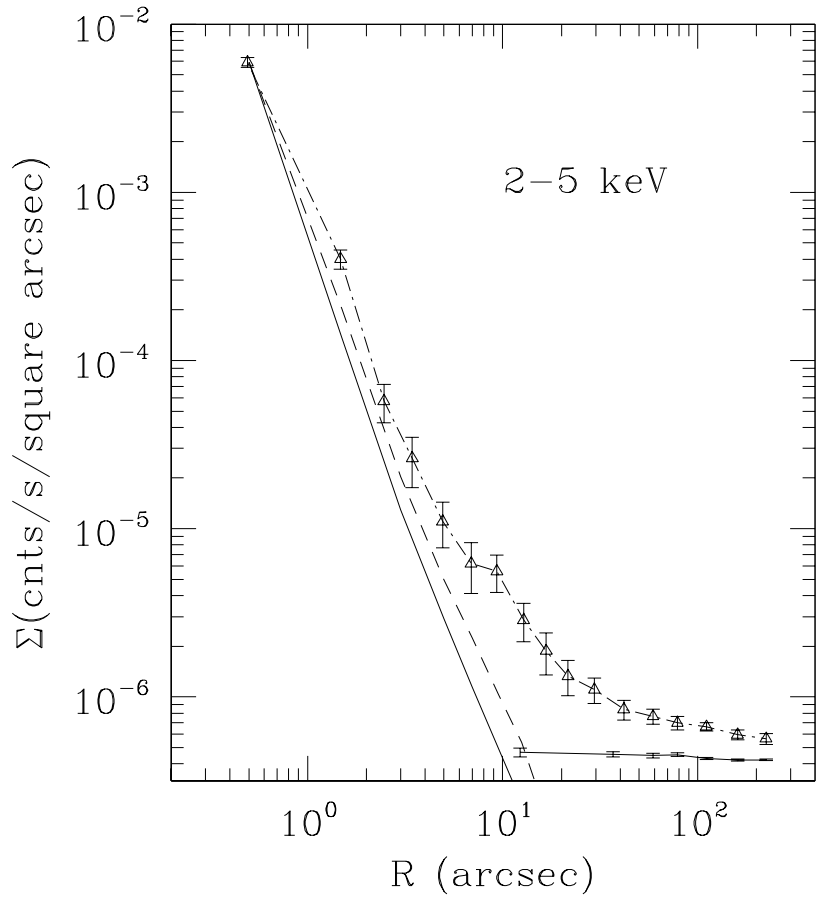}
\caption{The same as in Fig. 3 for the 2--5 keV band.
The PSF profile for photons of energy of 5 keV (dashed line) is also 
shown.  \label{fig4}}
\end{figure}

\clearpage

\begin{figure}
\epsscale{0.70}
\plotone{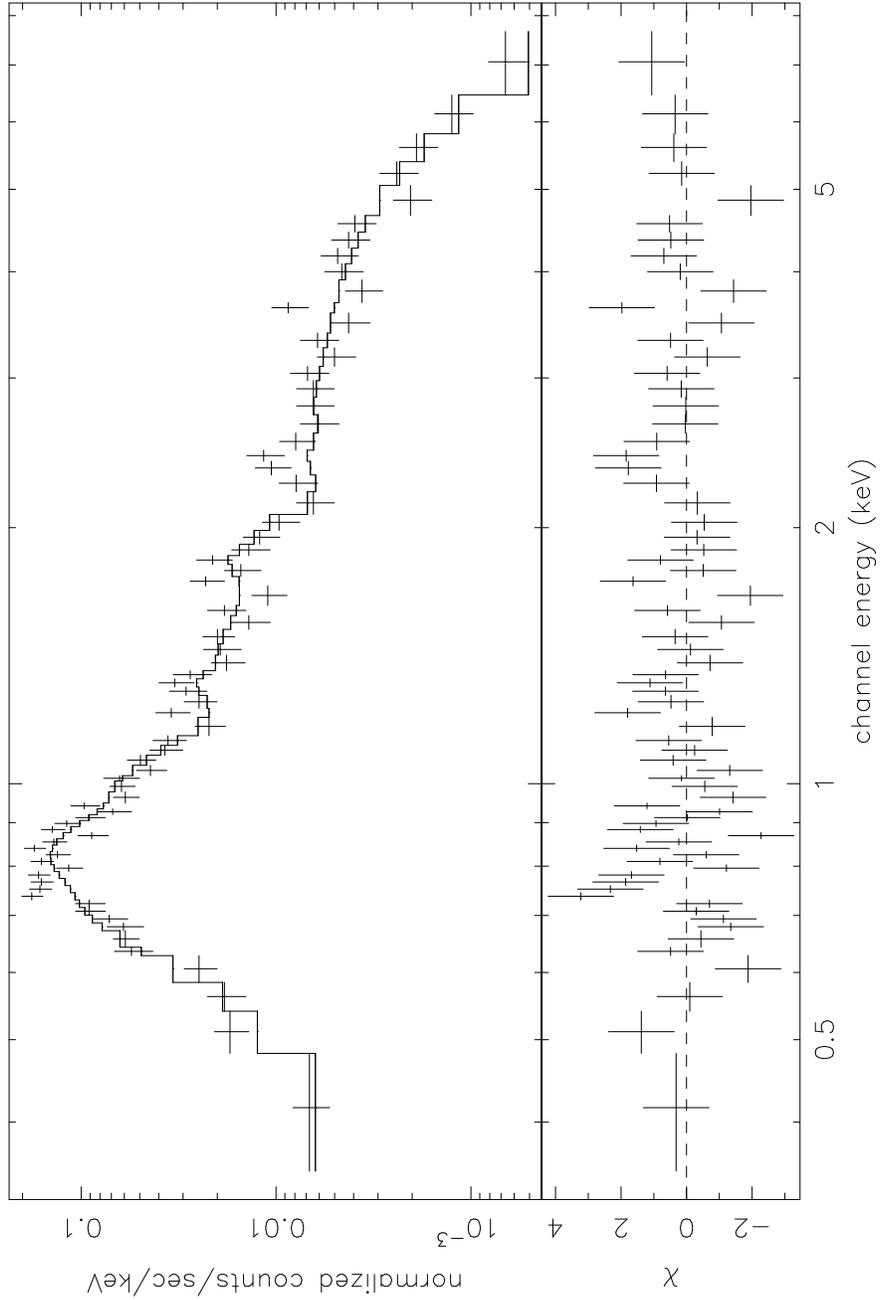}
\caption{The ACIS-S spectrum 
of the nuclear ($R<2^{\prime\prime}$) region of IC4296, together with
its modeling with a power law plus thermal emission (Sect. 3.2, Table 2). 
Folded spectra and residuals are shown.\label{fig5}}
\end{figure}

\clearpage

\begin{figure}
\epsscale{0.80}
\plotone{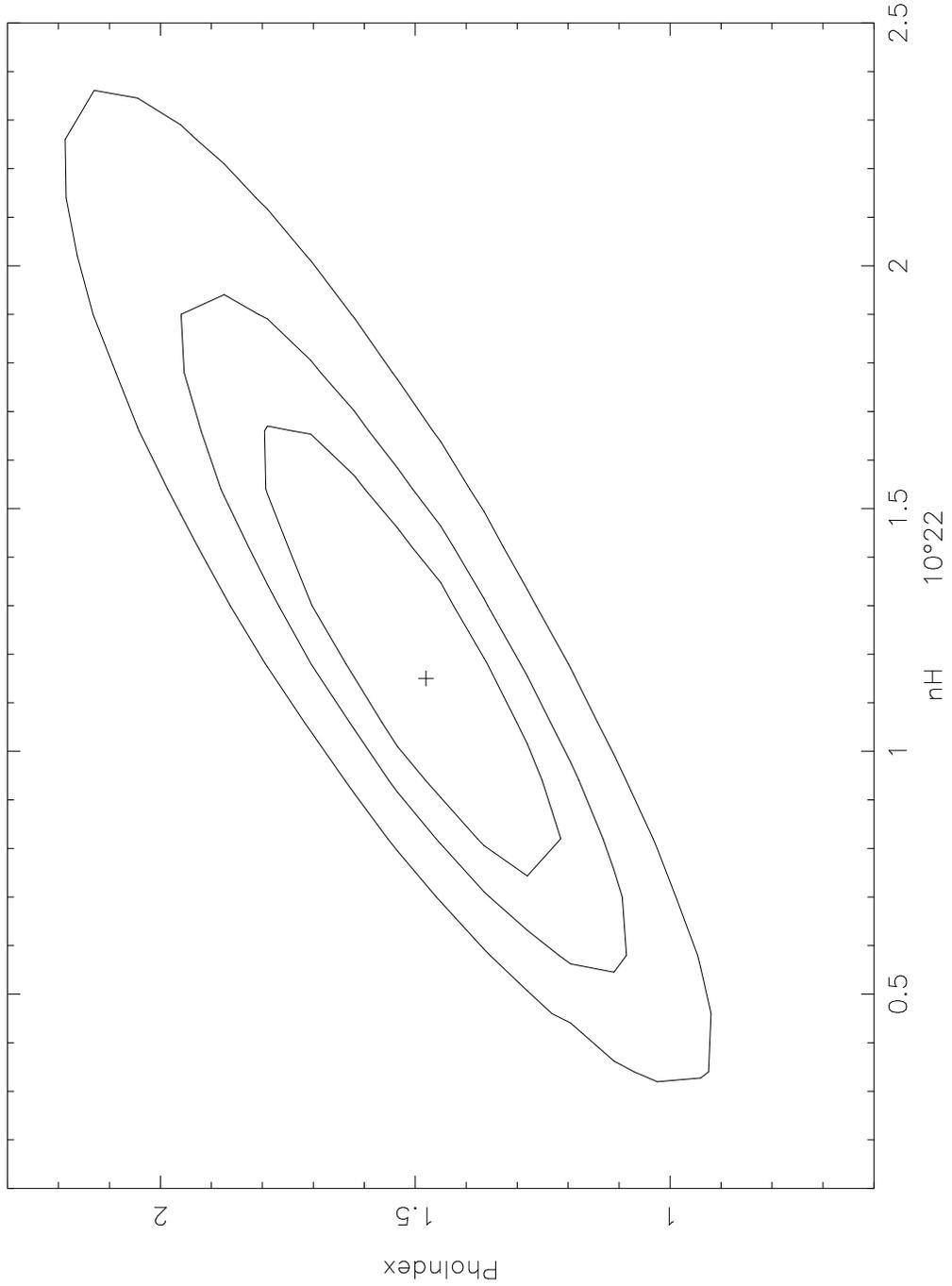}
 \caption{The 68\%, 90\% and 99\%
confidence contours for the photon index and the absorbing column
density of the power law component, in the modeling of the nuclear
spectrum with a power law plus thermal emission (Sect. 3.2, Table
2). \label{fig6}}
\end{figure}

\clearpage

\begin{figure}
\epsscale{0.80}
\plotone{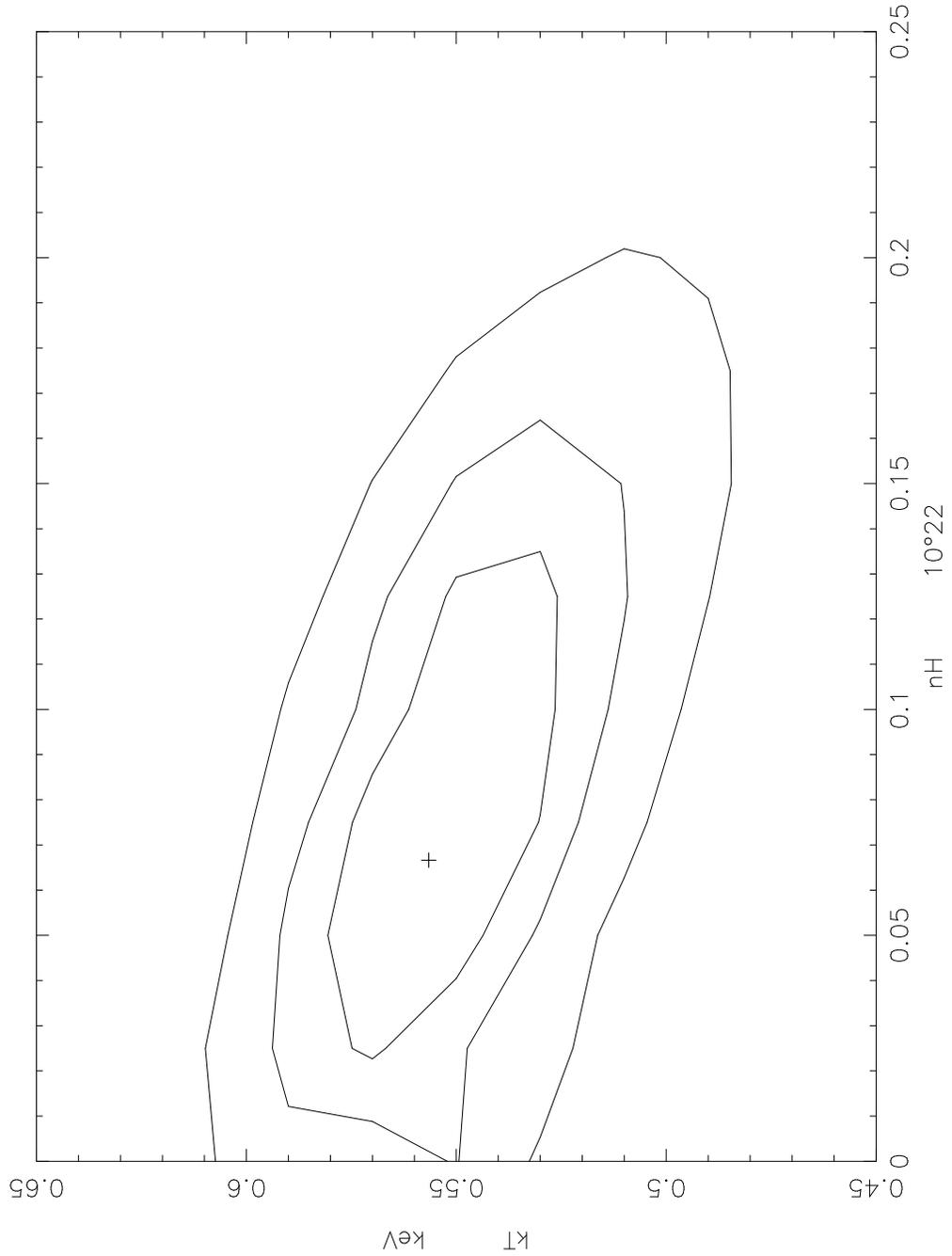}
\caption{The 68\%, 90\% and 99\%
confidence contours for the temperature and the absorbing column
density of the thermal component, in the modeling of the nuclear
spectrum with a power law plus thermal emission (Sect. 3.2, Table
2).\label{fig7} }
\end{figure}

\clearpage

\begin{figure}
\epsscale{0.90} 
\plotone{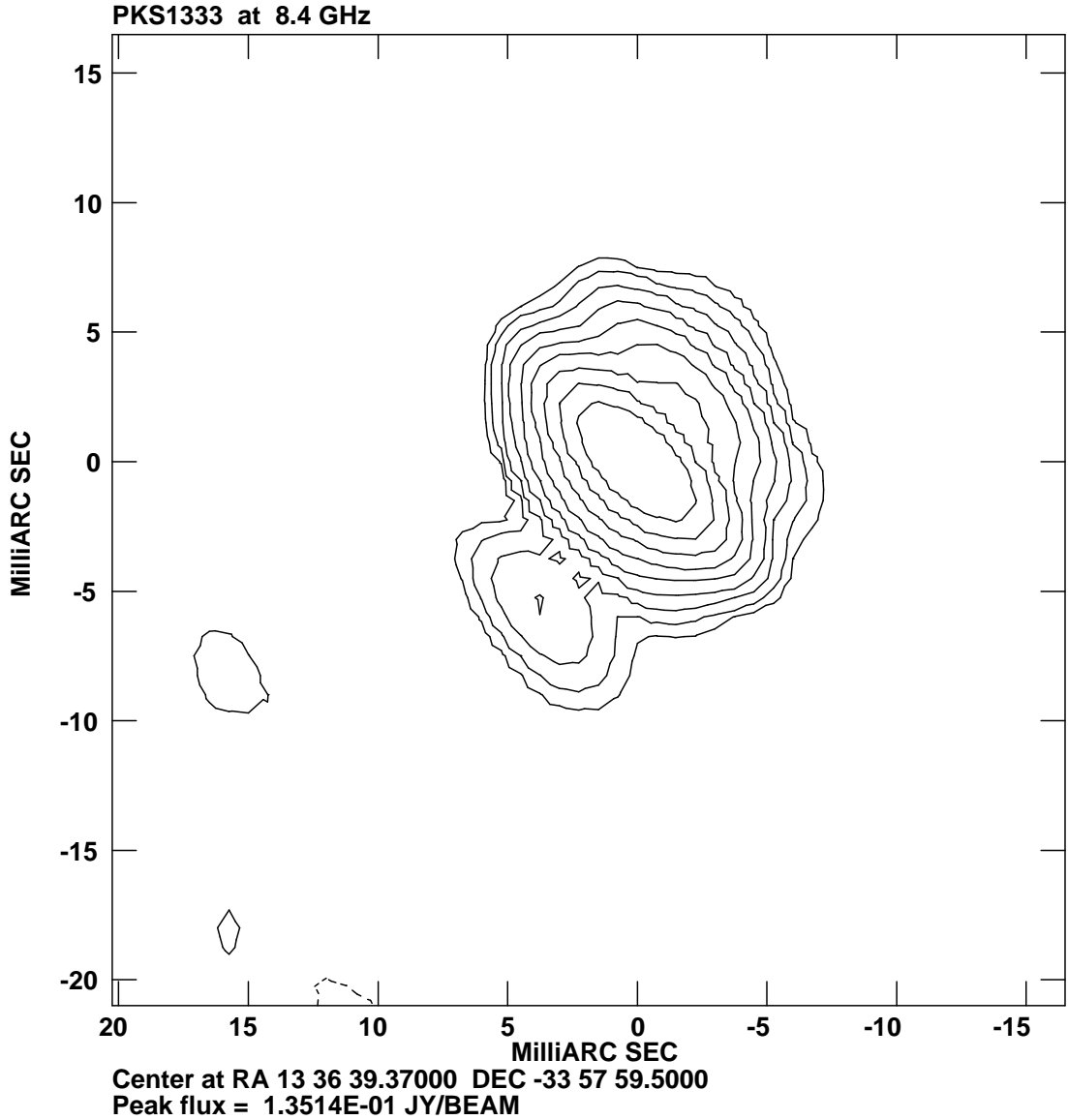}
\caption{
8.4 GHz VLBA image of PKS\,1333--33 (Sect. 4). The FWHM of the
restoring beam is 5$\times$2 mas at a position angle of 45$^{\circ}$. 
Contour levels are: --0.2, 0.2, 0.4, 0.8, 1.6, 3.2, 6.4, 12.8, 25, 50 
mJy/beam. The rms is 68 $\mu$Jy/beam.\label{fig8} }
\end{figure}

\clearpage

\begin{figure}
\plotone{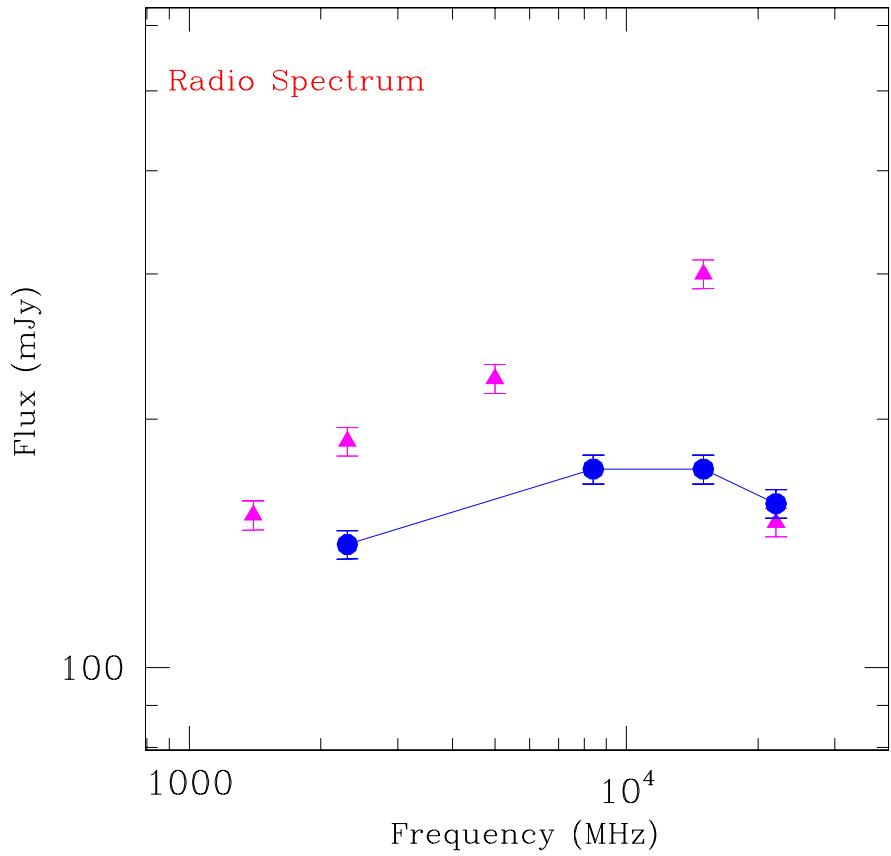}
\caption{Radio spectrum of PKS\,1333--33. Triangles 
refer to the kiloparsec scale core  (data from Killeen et al. 1986, 
Morganti et al. 1997); circles refer to the parsec scale
emission (see Sect. 4 for details). 
\label{fig9}}
\end{figure}

\clearpage 

\begin{figure}
\epsscale{1.0}
\plotone{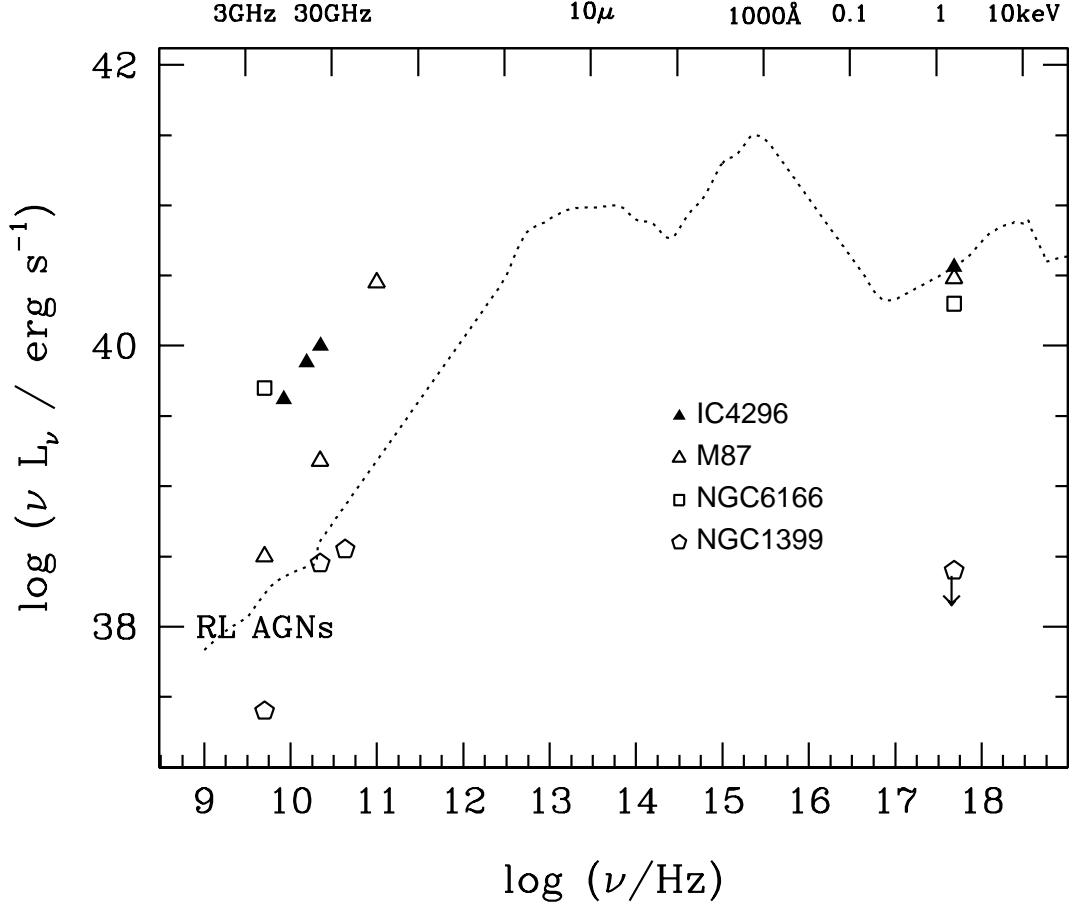}
\caption{Observed nuclear radio and X-ray luminosities
for IC4296 and for a few ellipticals for which an ADAF
modeling has been attempted (Sect. 5.2); the spectral energy distribution
of radio loud AGNs, normalized to the 2 keV luminosity
of IC4296, is also shown. X-ray luminosities derive from $Chandra$
observations and radio luminosities from VLBI or VLBA measurements
(except for NGC1399, for which there are only VLA measurements). The data
for IC4296 are given in this work, those for the other ellipticals
come from the references given in Sect. 5.2.
\label{fig10}}
\end{figure}

\end{document}